\documentclass[12pt,a4paper]{article}
\pdfoutput=1

\usepackage{ifthen} 
\newboolean{pdflatex}
\setboolean{pdflatex}{true} 

\newboolean{articletitles}
\setboolean{articletitles}{true} 

\newboolean{uprightparticles}
\setboolean{uprightparticles}{false} 

\newboolean{inbibliography}
\setboolean{inbibliography}{false} 

\usepackage{microtype}
\usepackage{lineno}  
\usepackage{xspace} 
\usepackage{caption} 

\usepackage{graphicx}  
\usepackage{color}
\usepackage{colortbl}
\graphicspath{{./figs/}} 

\usepackage{tabularx}
\usepackage{dcolumn}
\newcolumntype{d}[1]{D{.}{.}{#1}}

\usepackage{amsmath} 
\usepackage{amssymb}
\usepackage{amsfonts}
\usepackage{upgreek} 

\newcommand*\patchAmsMathEnvironmentForLineno[1]{%
\expandafter\let\csname old#1\expandafter\endcsname\csname #1\endcsname
\expandafter\let\csname oldend#1\expandafter\endcsname\csname
end#1\endcsname
 \renewenvironment{#1}%
   {\linenomath\csname old#1\endcsname}%
   {\csname oldend#1\endcsname\endlinenomath}%
}
\newcommand*\patchBothAmsMathEnvironmentsForLineno[1]{%
  \patchAmsMathEnvironmentForLineno{#1}%
  \patchAmsMathEnvironmentForLineno{#1*}%
}
\AtBeginDocument{%
\patchBothAmsMathEnvironmentsForLineno{equation}%
\patchBothAmsMathEnvironmentsForLineno{align}%
\patchBothAmsMathEnvironmentsForLineno{flalign}%
\patchBothAmsMathEnvironmentsForLineno{alignat}%
\patchBothAmsMathEnvironmentsForLineno{gather}%
\patchBothAmsMathEnvironmentsForLineno{multline}%
}

\usepackage{hyperref}    
\usepackage[all]{hypcap} 

\def\lhcb {\mbox{LHCb}\xspace}

\ifthenelse{\boolean{uprightparticles}}%
{

 \def\Ppi         {\ensuremath{\uppi}\xspace}

 \def\PDelta      {\ensuremath{\Delta}\xspace}                 
 \def\PXi      {\ensuremath{\Xi}\xspace}                 
 \def\PLambda      {\ensuremath{\Lambda}\xspace}                 
 \def\PSigma      {\ensuremath{\Sigma}\xspace}                 
 \def\POmega      {\ensuremath{\Omega}\xspace}                 
 \def\PUpsilon      {\ensuremath{\Upsilon}\xspace}                 
 
 \def\PB      {\ensuremath{\mathrm{B}}\xspace}                 
                  
 \def\PD      {\ensuremath{\mathrm{D}}\xspace}

 \def\PK      {\ensuremath{\mathrm{K}}\xspace}

 \def\Pb      {\ensuremath{\mathrm{b}}\xspace}                 
 \def\Pc      {\ensuremath{\mathrm{c}}\xspace}                 
 \def\Pd      {\ensuremath{\mathrm{d}}\xspace}

 \def\Pi      {\ensuremath{\mathrm{i}}\xspace}

 \def\Pp      {\ensuremath{\mathrm{p}}\xspace}

 \def\Ps      {\ensuremath{\mathrm{s}}\xspace}                 
                  
 \def\Pu      {\ensuremath{\mathrm{u}}\xspace}

}
{

 \def\Ppi         {\ensuremath{\pi}\xspace}

 \mathchardef\PDelta="7101
 \mathchardef\PXi="7104
 \mathchardef\PLambda="7103
 \mathchardef\PSigma="7106
 \mathchardef\POmega="710A
 \mathchardef\PUpsilon="7107
                  
 \def\PB      {\ensuremath{B}\xspace}                 
                  
 \def\PD      {\ensuremath{D}\xspace}

 \def\PK      {\ensuremath{K}\xspace}

 \def\Pb      {\ensuremath{b}\xspace}                 
 \def\Pc      {\ensuremath{c}\xspace}                 
 \def\Pd      {\ensuremath{d}\xspace}

 \def\Pi      {\ensuremath{i}\xspace}

 \def\Pp      {\ensuremath{p}\xspace}

 \def\Ps      {\ensuremath{s}\xspace}                 
                  
 \def\Pu      {\ensuremath{u}\xspace}

}

\def\uquark    {\ensuremath{\Pu}\xspace}

\def\dquark    {\ensuremath{\Pd}\xspace}

\def\squark    {\ensuremath{\Ps}\xspace}

\def\cquark    {\ensuremath{\Pc}\xspace}

\def\bquark    {\ensuremath{\Pb}\xspace}

\def\pion  {\ensuremath{\Ppi}\xspace}

\def\pip   {\ensuremath{\pion^+}\xspace}
\def\pim   {\ensuremath{\pion^-}\xspace}

\def\kaon  {\ensuremath{\PK}\xspace}
  \def\Kbar  {\kern 0.2em\overline{\kern -0.2em \PK}{}\xspace}

\def\Km    {\ensuremath{\kaon^-}\xspace}

  \def\Dbar    {\kern 0.2em\overline{\kern -0.2em \PD}{}\xspace}
\def\D       {\ensuremath{\PD}\xspace}

\def\Dz      {\ensuremath{\D^0}\xspace}

\def\Dp      {\ensuremath{\D^+}\xspace}

\def\Dstarp  {\ensuremath{\D^{*+}}\xspace}

\def\B       {\ensuremath{\PB}\xspace}
\def\Bbar    {\ensuremath{\kern 0.18em\overline{\kern -0.18em \PB}{}}\xspace}

\def\Bs      {\ensuremath{\B^0_\squark}\xspace}

  \def\Y#1S{\ensuremath{\PUpsilon{(#1S)}}\xspace}

\def\proton      {\ensuremath{\Pp}\xspace}

\def\Lbar {\ensuremath{\kern 0.1em\overline{\kern -0.1em\PLambda}}\xspace}
\def\Lambdares {\ensuremath{\PLambda}\xspace}

\def\Lc      {\ensuremath{\Lz^+_\cquark}\xspace}

\def\BF         {{\ensuremath{\cal B}\xspace}}

\newcommand{\decay}[2]{\ensuremath{#1\!\to #2}\xspace}         

\def\to                 {\ensuremath{\rightarrow}\xspace}

\def\AT#1     {\ensuremath{A_{\mathrm{T}}^{#1}}\xspace}           

\def\C#1      {\ensuremath{\mathcal{C}_{#1}}\xspace}                       
\def\Cp#1     {\ensuremath{\mathcal{C}_{#1}^{'}}\xspace}                    
\def\Ceff#1   {\ensuremath{\mathcal{C}_{#1}^{\mathrm{(eff)}}}\xspace}        
\def\Cpeff#1  {\ensuremath{\mathcal{C}_{#1}^{'\mathrm{(eff)}}}\xspace}       
\def\Ope#1    {\ensuremath{\mathcal{O}_{#1}}\xspace}                       
\def\Opep#1   {\ensuremath{\mathcal{O}_{#1}^{'}}\xspace}                    

\newcommand{\tev}{\ifthenelse{\boolean{inbibliography}}{\ensuremath{~T\kern -0.05em eV}\xspace}{\ensuremath{\mathrm{\,Te\kern -0.1em V}}\xspace}}
\newcommand{\gev}{\ensuremath{\mathrm{\,Ge\kern -0.1em V}}\xspace}
\newcommand{\mev}{\ensuremath{\mathrm{\,Me\kern -0.1em V}}\xspace}
\newcommand{\kev}{\ensuremath{\mathrm{\,ke\kern -0.1em V}}\xspace}
\newcommand{\ev}{\ensuremath{\mathrm{\,e\kern -0.1em V}}\xspace}
\newcommand{\gevc}{\ensuremath{{\mathrm{\,Ge\kern -0.1em V\!/}c}}\xspace}
\newcommand{\mevc}{\ensuremath{{\mathrm{\,Me\kern -0.1em V\!/}c}}\xspace}
\newcommand{\gevcc}{\ensuremath{{\mathrm{\,Ge\kern -0.1em V\!/}c^2}}\xspace}
\newcommand{\gevgevcccc}{\ensuremath{{\mathrm{\,Ge\kern -0.1em V^2\!/}c^4}}\xspace}
\newcommand{\mevcc}{\ensuremath{{\mathrm{\,Me\kern -0.1em V\!/}c^2}}\xspace}

\def\mum  {\ensuremath{\,\upmu\rm m}\xspace}

\def\mub{\ensuremath{\rm \,\upmu b}\xspace}
\def\nb {\ensuremath{\rm \,nb}\xspace}

\def\invfb   {\ensuremath{\mbox{\,fb}^{-1}}\xspace}

\def\fs   {\ensuremath{\rm \,fs}\xspace}

\newcommand{\chisq}{\ensuremath{\chi^2}\xspace}

\newcommand{\chisqip}{\ensuremath{\chi^2_{\rm IP}}\xspace}

\def\gsim{{~\raise.15em\hbox{$>$}\kern-.85em
          \lower.35em\hbox{$\sim$}~}\xspace}
\def\lsim{{~\raise.15em\hbox{$<$}\kern-.85em
          \lower.35em\hbox{$\sim$}~}\xspace}

\def\pt         {\mbox{$p_{\rm T}$}\xspace}
\def\et         {\mbox{$E_{\rm T}$}\xspace}

\def\evtgen     {\mbox{\textsc{EvtGen}}\xspace}

\def\geant      {\mbox{\textsc{Geant4}}\xspace}

\def\photos     {\mbox{\textsc{Photos}}\xspace}

\def\pythia     {\mbox{\textsc{Pythia}}\xspace}

\def\tell1  {TELL1\xspace}
\def\ukl1   {UKL1\xspace}

\newcommand{\eg}{\mbox{\itshape e.g.}\xspace}

\usepackage{cite} 
\usepackage{mciteplus}
\usepackage{float}

\def\Lc      {\ensuremath{\Lambdares_\cquark}\xspace}

\def\Lcp     {\ensuremath{\Lambdares_\cquark^+}\xspace}

\def\Xiccbare          {\ensuremath{\Xi_{\cquark\cquark}}\xspace}
\def\Xicc              {\ensuremath{\Xi_{\cquark\cquark}^+}\xspace}

\def\Xiccp             {\Xicc}

\def\mLcp         {\ensuremath{m([\proton \Km \pip]_{\Lc})}\xspace}

\def\mXicc        {\ensuremath{m([\proton \Km \pip]_{\Lc} \Km \pip)}\xspace}

\def\XiccpToLcpKmpip     {\mbox{\ensuremath{\decay{\Xiccp}{\Lcp\Km\pip}}}\xspace}

\def\LcpTopKmpip        {\mbox{\ensuremath{\decay{\Lcp}{\proton\Km\pip}}}\xspace}  

\def\genxicc      {\mbox{\textsc{Genxicc}}\xspace}

\def\pvalue      {\mbox{$p$-value}\xspace}
\def\pvalues     {\mbox{$p$-values}\xspace}

\usepackage[mathscr]{euscript} 

\begin{document}

\renewcommand{\thefootnote}{\fnsymbol{footnote}}
\setcounter{footnote}{1}


\begin{titlepage}
\pagenumbering{roman}

\vspace*{-1.5cm}
\centerline{\large EUROPEAN ORGANIZATION FOR NUCLEAR RESEARCH (CERN)}
\vspace*{1.5cm}
\hspace*{-0.5cm}
\begin{tabular*}{\linewidth}{lc@{\extracolsep{\fill}}r}
\vspace*{-2.7cm}\mbox{\!\!\!\includegraphics[width=.14\textwidth]{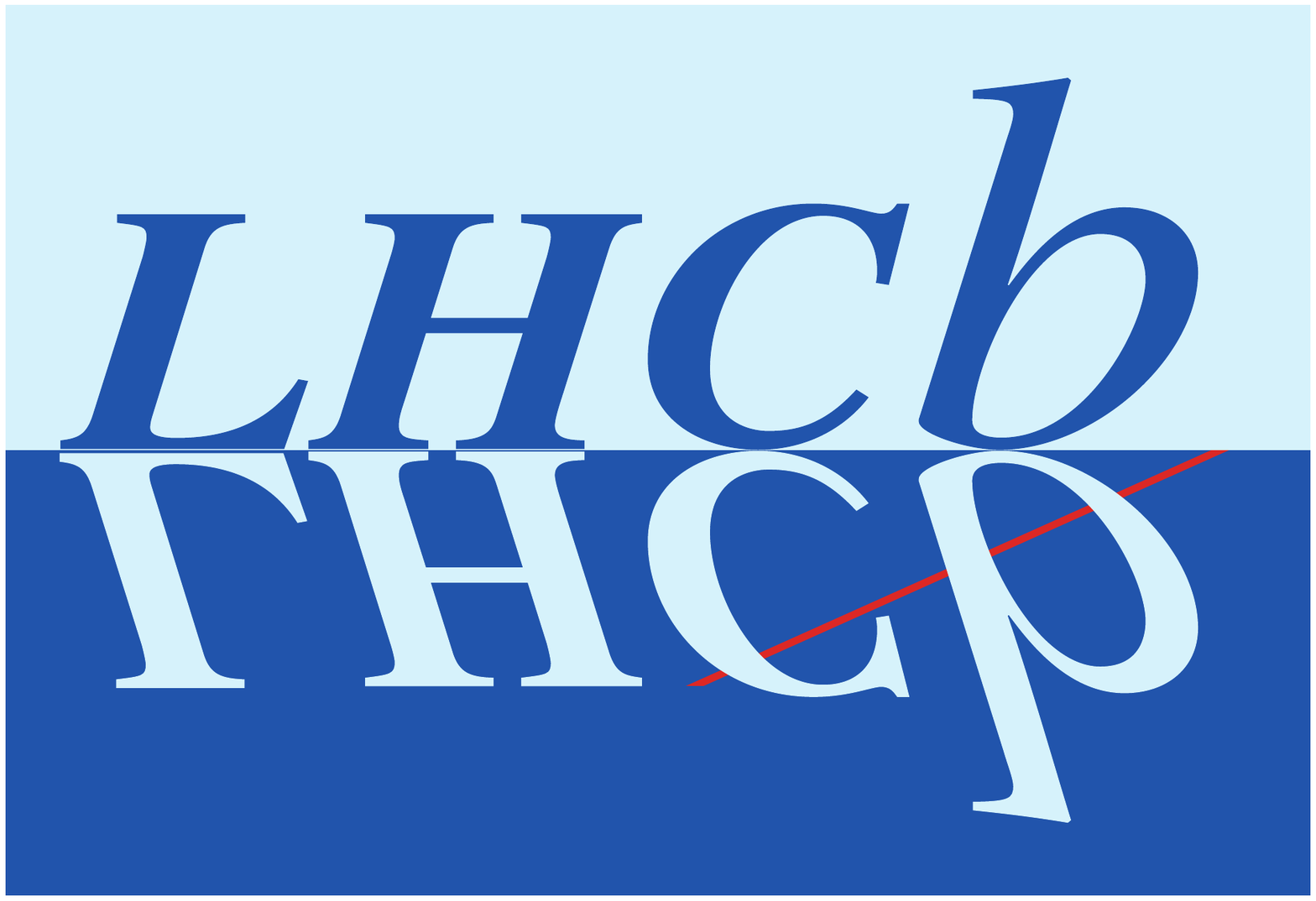}} & &
\\
 & & CERN-PH-EP-2013-181 \\  
 & & LHCb-PAPER-2013-049 \\  
 & & January 6, 2014 \\ 
 & & \\
\end{tabular*}

\vspace*{4.0cm}

{\bf\boldmath\huge
\begin{center}
  Search for the doubly charmed baryon \Xiccp
\end{center}
}

\vspace*{2.0cm}

\begin{center}
The LHCb collaboration\footnote{Authors are listed on the following pages.}
\end{center}

\vspace{\fill}

\begin{abstract}
  \noindent
  A search for the doubly charmed baryon \Xiccp in the decay mode
  \XiccpToLcpKmpip is performed 
  with a data sample,
  corresponding to an integrated luminosity of 0.65\invfb,
  of $\proton\proton$ collisions
  recorded at a centre-of-mass energy of 7\tev.
  No significant signal is found in the 
  mass range 3300--3800\mevcc.
  Upper limits at the 95\% confidence level on the ratio of the \Xiccp production cross-section times
  branching fraction
  to that of the \Lcp, $R$, are given as a function of the \Xiccp mass and lifetime.
  The largest upper limits range from
  $R<1.5 \times 10^{-2}$ for a lifetime of 100\fs to
  $R<3.9 \times 10^{-4}$ for a lifetime of 400\fs.
\end{abstract}

\vspace*{2.0cm}

\begin{center}
  Published in JHEP, DOI: \href{http://dx.doi.org/10.1007/JHEP12(2013)090}{10.1007/JHEP12(2013)090}
\end{center}

\vspace{\fill}

{\footnotesize 
\centerline{\copyright~CERN on behalf of the \lhcb collaboration, license \href{http://creativecommons.org/licenses/by/3.0/}{CC-BY-3.0}.}}
\vspace*{2mm}

\end{titlepage}


\newpage
\setcounter{page}{2}
\mbox{~}
\newpage

\centerline{\large\bf LHCb collaboration}
\begin{flushleft}
\small
R.~Aaij$^{40}$, 
B.~Adeva$^{36}$, 
M.~Adinolfi$^{45}$, 
C.~Adrover$^{6}$, 
A.~Affolder$^{51}$, 
Z.~Ajaltouni$^{5}$, 
J.~Albrecht$^{9}$, 
F.~Alessio$^{37}$, 
M.~Alexander$^{50}$, 
S.~Ali$^{40}$
% ...

{\footnotesize \it
% ...
$ ^{5}$Clermont Universit\'{e}, Universit\'{e} Blaise Pascal, CNRS/IN2P3, LPC, Clermont-Ferrand, France\\
$ ^{6}$CPPM, Aix-Marseille Universit\'{e}, CNRS/IN2P3, Marseille, France\\
% ...
$ ^{9}$Fakult\"{a}t Physik, Technische Universit\"{a}t Dortmund, Dortmund, Germany\\
% ...
$ ^{36}$Universidad de Santiago de Compostela, Santiago de Compostela, Spain\\
$ ^{37}$European Organization for Nuclear Research (CERN), Geneva, Switzerland\\
% ...
$ ^{40}$Nikhef National Institute for Subatomic Physics, Amsterdam, The Netherlands\\
% ...
$ ^{45}$H.H. Wills Physics Laboratory, University of Bristol, Bristol, United Kingdom\\
% ...
$ ^{50}$School of Physics and Astronomy, University of Glasgow, Glasgow, United Kingdom\\
$ ^{51}$Oliver Lodge Laboratory, University of Liverpool, Liverpool, United Kingdom\\
% ...
}
\end{flushleft}



\cleardoublepage


\renewcommand{\thefootnote}{\arabic{footnote}}
\setcounter{footnote}{0}



\pagestyle{plain} 
\setcounter{page}{1}
\pagenumbering{arabic}



\section{Introduction}
\label{sec:intro}

The constituent quark model~\cite{GellMann:1964nj,Zweig:1981pd,Zweig:1964jf}
predicts the existence of multiplets of baryon
and meson states, with a structure determined
by the symmetry properties of the hadron
wavefunctions. When considering \uquark, \dquark, \squark, and \cquark
quarks, the states form $SU(4)$ multiplets~\cite{DeRujula:1975ge}.
The baryon ground states---those with no orbital
or radial excitations---consist of a 20-plet with
spin-parity $J^P = 1/2^+$ and a 20-plet with $J^P = 3/2^+$.
All of the ground states with charm quantum number $C=0$ or $C=1$ have
been discovered~\cite{PDG2012}.
Three weakly decaying $C=2$ states are expected:
a \Xiccbare isodoublet ($\cquark\cquark\uquark, \cquark\cquark\dquark$) and
an $\Omega_{\cquark\cquark}$ isosinglet ($\cquark\cquark\squark$), each with
$J^P = 1/2^+$. 
This paper reports a search for the
\Xiccp baryon.
There are numerous predictions for
the masses of these states
(see, \eg, Ref.~\cite{Roberts:2007ni} and the references therein,
as well as Refs.~\cite{He:2004px,wang2010analysis, Chang:2006eu, Valcarce:2008dr,Zhang:2008rt})
with most estimates for the \Xiccp mass
in the range 3500--3700\mevcc.
Predictions for its lifetime
range between 100 and 250\fs~\cite{hsi2008lifetime, Ebert:2002ig, guberina1999inclusive}.

Signals for the \Xiccp baryon were reported in the $\Lcp \Km \pip$ and 
$\proton \Dp \Km$ final states by the SELEX collaboration, 
using a hyperon beam
(containing an admixture of \proton, $\Sigma^-$, and \pim)
on a fixed target ~\cite{Mattson:2002vu,Ocherashvili:2004hi}.
The mass was measured to be $3519 \pm 2$\mevcc, and the lifetime was
found to be compatible with zero within
experimental resolution and less than 33\fs at the 90\% confidence level~(CL).
SELEX estimated that 20\% of their \Lcp yield
originates from \Xiccp decays,
in contrast to theory expectations that the
production of doubly charmed baryons would be suppressed
by several orders of magnitude with respect to singly charmed baryons~\cite{kiselev2002baryons}.
Searches in different production environments at the FOCUS, BaBar, and Belle experiments have not
shown evidence for a \Xiccp state with the properties reported by
SELEX~\cite{ratti2003new,Aubert:2006qw,Chistov:2006zj}.

This paper presents the result
of a search for the
decay\footnote{The inclusion of charge-conjugate processes is implied throughout.}
\XiccpToLcpKmpip
with the LHCb detector and an integrated luminosity of $0.65\invfb$ of $pp$ collision data recorded at
centre-of-mass energy $\sqrt{s}=7\tev$.
Double charm production has been observed previously at LHCb
both in the $J/\psi \, J/\psi$ final state~\cite{LHCb-PAPER-2011-013} and in final states including
one or two open charm hadrons~\cite{LHCb-PAPER-2012-003}.
Phenomenological estimates of the production cross-section of
\Xiccbare in $\proton\proton$ collisions at $\sqrt{s}=14\tev$ are
in the range 60--1800\nb~\cite{kiselev2002baryons,Ma:2003zk,Chang:2006xp}; the
cross-section at $\sqrt{s}=7\tev$ is expected to be roughly a factor of two smaller. As is
typical for charmed hadrons, the production is expected to be concentrated
in the low transverse momentum (\pt) 
and forward rapidity ($y$) kinematic region instrumented by LHCb~\cite{Chang:2006xp}.
For comparison, the prompt \Lcp cross-section in
the range $0 < \pt < 8000$\mevc and $2.0 < y < 4.5$ at $\sqrt{s}=7\tev$ has been
measured to be $(233 \pm 26 \pm 71 \pm 14)$\mub at LHCb~\cite{LHCB-PAPER-2012-041}, where the uncertainties are
statistical, systematic, and due to the description of the fragmentation model,
respectively.
Thus, the cross-section for \Xiccp production at LHCb is predicted to be
smaller than that for \Lcp by a factor of order $10^{-4}$ to $10^{-3}$.

To reduce systematic uncertainties, 
the \Xiccp cross-section is measured
relative to that of the \Lcp. This has the further advantage that it
allows a direct comparison with previous experimental results.
The production ratio $R$ that is measured is defined as
\begin{equation}
R \equiv \frac{\sigma(\Xicc) \,  \BF(\XiccpToLcpKmpip)}{\sigma(\Lcp) }
=
\frac{ 
  N_{\text{sig}}
}{
  N_{\text{norm}}
}
\frac{ 
  \varepsilon_{\text{norm}}
}{
  \varepsilon_{\text{sig}}
} ,
\label{eq:defineMainResult}
\end{equation}
where 
 $N_{\text{sig}}$ and $N_{\text{norm}}$ refer to the measured yields of the 
 signal (\Xiccp) and normalisation (\Lcp) modes,
 $\varepsilon_{\text{sig}}$ and $\varepsilon_{\text{norm}}$ are the corresponding efficiencies, 
 $\BF$ indicates a branching fraction, and
 $\sigma$ indicates a cross-section.
Assuming that 
  $\BF(\XiccpToLcpKmpip) \approx
  \BF(\LcpTopKmpip) \approx 5\%$~\cite{PDG2012},
the expected value of $R$ at LHCb is of order $10^{-5}$ to $10^{-4}$.
  By contrast, the SELEX observation~\cite{Mattson:2002vu}
  reported 15.9~\Xiccp signal events in a sample of
  1630~\Lcp events with an efficiency ratio of 11\%,
  corresponding to $R=9\%$.
For convenience, the single-event sensitivity $\alpha$ is defined as
\begin{equation}
  \alpha \equiv \frac{
    \varepsilon_{\text{norm}}
  }{
    N_{\text{norm}} \, \varepsilon_{\text{sig}}
  }
  \label{eq:defAlpha}
\end{equation}
such that $R = \alpha N_{\text{sig}}$. 
For each candidate the mass difference $\delta m$ is computed as
\begin{equation}
  \delta m \equiv \mXicc - \mLcp - m(\Km) - m(\pip)
  ,
  \label{eq:defineDeltaM}
\end{equation}
where \mXicc is the measured invariant mass of the \Xiccp candidate,
\mLcp is the measured invariant mass of the $\proton \Km \pip$ combination forming the \Lcp candidate, and
$m(\Km)$ and $m(\pip)$ are the world-average masses of
charged kaons and pions, respectively~\cite{PDG2012}.

Since no assumption is made about the \Xiccp mass, a wide signal window of
$380 < \delta m < 880$\mevcc is used for this search, corresponding
to approximately $3300 < m(\Xiccp) < 3800$\mevcc.
All aspects of the analysis procedure were fixed before the data
in this signal region were examined.
Limits on $R$ are quoted as a function of the \Xiccp
mass and lifetime, since the measured yield depends on
$\delta m$, and $\varepsilon_{\text{sig}}$ depends on both the mass and lifetime.

\section{Detector and software}
\label{sec:Detector}

The \lhcb detector~\cite{Alves:2008zz} is a single-arm forward
spectrometer covering the \mbox{pseudorapidity} range $2<\eta <5$,
designed for the study of particles containing \bquark or \cquark
quarks. The detector includes a high-precision tracking system
consisting of a silicon-strip vertex detector (VELO) surrounding the $pp$
interaction region, a large-area silicon-strip detector located
upstream of a dipole magnet with a bending power of about
$4{\rm\,Tm}$, and three stations of silicon-strip detectors and straw
drift tubes placed downstream.
The combined tracking system provides a momentum measurement with
relative uncertainty that varies from 0.4\% at 5\gevc to 0.6\% at 100\gevc,
and impact parameter (IP) resolution of 20\mum for
tracks with large transverse momentum. Charged hadrons are identified
using two ring-imaging Cherenkov detectors~\cite{LHCb-DP-2012-003}. Photon, electron, and
hadron candidates are identified by a calorimeter system consisting of
scintillating-pad and preshower detectors, an electromagnetic
calorimeter, and a hadronic calorimeter. Muons are identified by a
system composed of alternating layers of iron and multiwire
proportional chambers~\cite{LHCb-DP-2012-002}.
The trigger~\cite{LHCb-DP-2012-004} consists of a
hardware stage, based on information from the calorimeter and muon
systems, followed by a software stage, which applies a full event
reconstruction.

In the simulation, $pp$ collisions are generated using
\pythia~6.4~\cite{Sjostrand:2006za} with a specific \lhcb
configuration~\cite{LHCb-PROC-2010-056}.
A dedicated generator, \genxicc~v2.0, is used to
simulate \Xiccp baryon production~\cite{Chang:2009va}.
Decays of hadronic particles
are described by \evtgen~\cite{Lange:2001uf}, in which final state
radiation is generated using \photos~\cite{Golonka:2005pn}. The
interaction of the generated particles with the detector and its
response are implemented using the \geant
toolkit~\cite{Allison:2006ve, *Agostinelli:2002hh} as described in
Ref.~\cite{LHCb-PROC-2011-006}.
Unless otherwise stated, simulated events are generated
with $m(\Xiccp)=3500$\mevcc, with $\tau_{\Xiccp}=333$\fs,
and with the \Xiccp and \Lcp decay products distributed
according to phase space.

\section{Triggering, reconstruction, and selection}
\label{sec:sel}

The procedure to trigger, reconstruct,
and select candidates for the signal and normalisation
modes is designed to retain signal
and to suppress three  primary sources of background.
These are
  combinations of unrelated tracks, especially those
    originating from the same primary interaction vertex (PV);
  mis-reconstructed charm or beauty hadron decays, which
    typically occur at a displaced vertex;
  and combinations of a real \Lcp with other tracks to
    form a fake \Xiccp candidate.
The first two classes generally have a smooth distribution
in both \mLcp and $\delta m$; the third peaks in \mLcp
but is smooth in $\delta m$.

For both the \Xiccp search and
the normalisation mode, \Lcp candidates are reconstructed
in the final state $\proton \Km \pip$.
To minimise systematic differences in efficiency between
the signal and normalisation modes, the same trigger requirements
are used for both modes, and those requirements
ensure that the event was triggered by the
\Lcp candidate and its daughter tracks.
  First, at least
one of the three \Lcp daughter tracks must
correspond to a calorimeter cluster with a measured
transverse energy $\et > 3500$\mev in the hardware trigger.
  Second, at least one of the three
\Lcp daughter tracks must be selected by the inclusive
software trigger, which requires that the track have
$\pt > 1700\mevc$ and $\chisqip > 16$ with respect to any
PV, where \chisqip is defined as the
difference in \chisq of a given PV reconstructed with and
without the considered track.
  Third, the \Lcp candidate must be reconstructed
and accepted by a dedicated \LcpTopKmpip selection
algorithm in the software trigger. This algorithm makes several
geometric and kinematic requirements, the most important of which are as follows.
The three daughter tracks are required
  to have $\pt > 500$\mevcc,
  to have a track fit $\chisq/\rm{ndf} < 3$,
  not to originate at a PV ($\chisqip > 16$),
  and to meet at a common vertex ($\chisq/\rm{ndf}<15$, where $\rm{ndf}$ is the number of degrees of freedom).
The \Lcp candidate formed from the three tracks is required
  to have $\pt > 2500$\mevcc,
  to lie within the mass window $2150 < \mLcp < 2430$\mevcc,
  to be significantly displaced from the PV (vertex separation $\chisq > 16$),
  and to point back towards the PV (momentum and displacement vectors within $1^\circ$).
The software trigger also requires that the proton candidate
be inconsistent with the pion and kaon mass hypotheses.
The \Lcp trigger algorithm was only enabled for part of the
data-taking in 2011, corresponding to an integrated luminosity of
$0.65\invfb$.

For events that pass the trigger,
the \Lcp selection proceeds in a similar fashion to
that used in the software trigger: three charged tracks are
required to form a common vertex that is significantly
displaced from the event PV and has invariant mass in the 
range $2185 < \mLcp < 2385$\mevcc.
Particle identification (PID) requirements are imposed on all three
tracks to suppress combinatorial background and
mis-identified charm meson decays. The same \Lcp selection
is used for the signal and normalisation modes.

The \Xiccp candidates are formed by combining a \Lcp
candidate with two tracks, one identified as a \Km and one as a \pip.
These three particles are required to form a common vertex ($\chisq/\rm{ndf}<10$)
that is displaced from the PV (vertex separation $\chisq > 16$).
The kaon and pion daughter tracks are also required to not originate at the
PV ($\chisqip > 16$) and to have $\pt > 250$\mevc.
The \Xiccp candidate is required to point back to the PV
and to have $\pt > 2000$\mevc.

A multivariate selection is applied only to the signal mode
  to further improve the purity.
The selector used is an artificial neural network (ANN)
implemented in the TMVA package~\cite{Hocker:2007ht}.
The input variables
are chosen to have limited dependence on 
the \Xiccp lifetime. To train the selector,
simulated \Xiccp decays are used as the signal sample and
$3.5\%$
of the candidates from $\delta m$ sidebands of width 200\mevcc
adjacent to the signal region are used as the background sample.
  In order to increase the available statistics, the trigger requirements
  are relaxed for these samples.
In addition to the training samples, disjoint test samples of equal
size are taken from the same sources.
After training, the response distribution of the ANN is compared
between the training and test samples. Good agreement is found for
both signal and background, with Kolmogorov-Smirnov
test \pvalues of 80\% and 65\%, respectively.
A selection cut on the ANN response is applied to the data used in
the \Xiccp search.
In the test samples, the efficiency of this requirement is
$55.7\%$ for signal and $4.2\%$ for background.

The selection has limited efficiency for
short-lived \Xiccp. This is principally due to the requirements
that the \Xiccp decay vertex be significantly displaced from the
PV, and that the \Xiccp daughter kaon and pion have a significant
impact parameter with respect to the PV.
As a consequence, the analysis is insensitive
to $\Xi_{\cquark}$ resonances that decay strongly to the same final state,
notably the $\Xi_{\cquark}(2980)^+$, $\Xi_{\cquark}(3055)^+$, and
$\Xi_{\cquark}(3080)^+$~\cite{Chistov:2006zj,Aubert:2007dt}.

\section{Yield measurements}
\label{sec:yield:sig}

\begin{figure}
  \begin{center}
    \includegraphics[width = 0.72\linewidth]{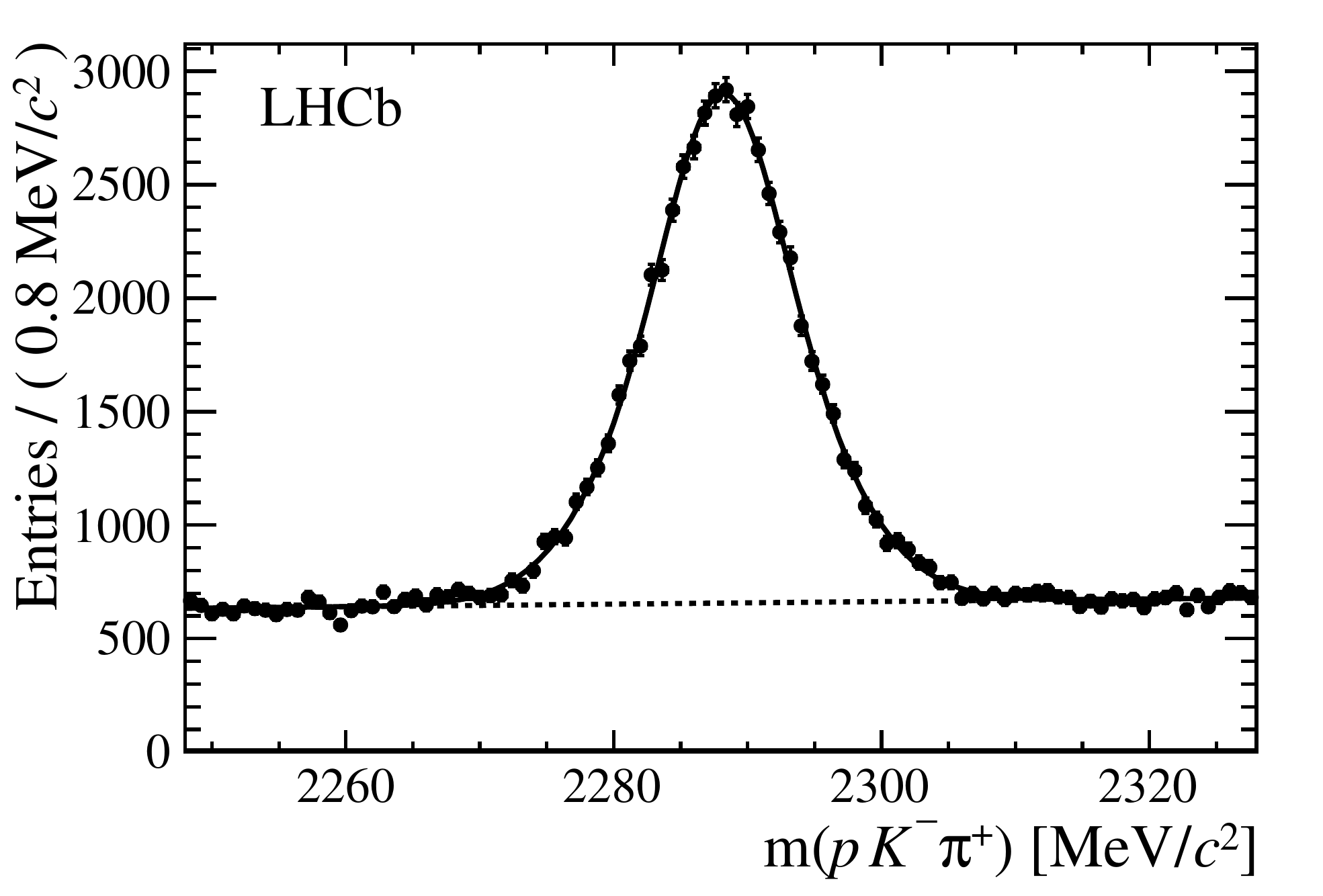}
  \end{center}
  \caption{
    \small
    Invariant mass spectrum of $\LcpTopKmpip$ candidates
    for 5$\%$ of the data, with events chosen at random
    during preselection
    (due to bandwidth limits for the normalisation mode).
    The dashed line shows the fitted
    background contribution, and the solid line the sum
    of \Lcp signal and background.
  }
  \label{fig:Lc}
\end{figure}

To determine the \Lcp yield, $N_{\text{norm}}$,
a fit is performed to the $\proton \Km \pip$ mass spectrum.
The signal shape is described as the sum of two
Gaussian functions with a common mean, and the background
is parameterised as a first-order polynomial.
The fit is shown in Fig.~\ref{fig:Lc}.
The selected \Lcp yield in the full $0.65\invfb$ sample is
$N_{\text{norm}} = (818 \pm 7) \times 10^3$, with
an invariant mass resolution of around
6\mevcc. 

\begin{figure}
  \begin{center}
    \includegraphics[width=0.72\linewidth]{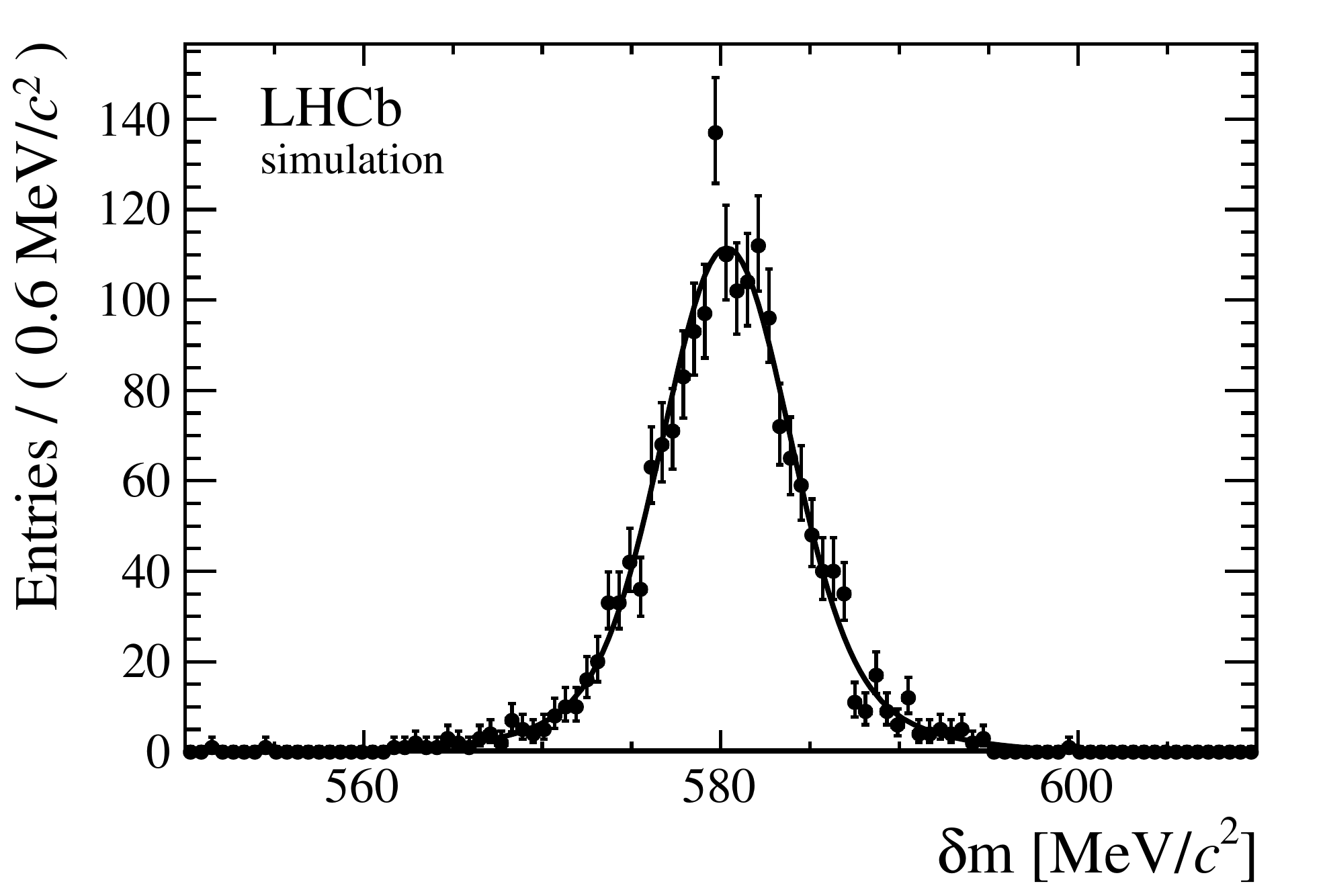}
  \end{center}
  \caption{
    \small
    The distribution of the invariant mass difference $\delta m$,
    defined in Eq.~\ref{eq:defineDeltaM},
    for simulated \Xiccp events with a \Xiccp mass of 3500\mevcc.
    The solid line shows the fitted signal shape.
      In order to increase the available statistics, the trigger and ANN
      requirements are not applied in this plot.
  }
  \label{fig:XiccMC}
\end{figure}

The \Xiccp signal yield is measured from the $\delta m$ distribution
under a series of different mass hypotheses. 
Although the methods used are designed not to require detailed
knowledge of the signal shape, it is necessary to know the
resolution with sufficient precision to define a signal window. Since the \Xiccp
yield may be small, its resolution cannot be
measured from data and is instead estimated with a sample of
simulated events, shown in Fig.~\ref{fig:XiccMC}.
Fitting the candidates with the sum of two Gaussian functions,
the resolution is found to be approximately 4.4\mevcc.

Two complementary procedures are used to estimate the signal
yield given a mass hypothesis $\delta m_0$. Both follow
the same general approach, but use different methods
to estimate the background. In both cases, a narrow signal window is
defined as $2273 < \mLcp < 2303$\mevcc and
$\left| \delta m - \delta m_0 \right| < 10$\mevcc, and the number of
candidates inside that window is taken as $N_{S+B}$.
Candidates outside the narrow window are used to estimate the expected background $N_B$
inside the window. The signal yield is then 
$N_S = N_{S+B} - N_B$. This avoids any need to model the signal
shape beyond an efficiency correction for the estimated
signal fraction lost outside
the window of width 20\mevcc.

The first method is an analytic, two-dimensional sideband
subtraction in \mLcp and $\delta m$. A two-dimensional region
of width 80\mevcc in \mLcp and width 200\mevcc in $\delta m$
is centred around the narrow signal window.
A $5 \times 5$ array of non-overlapping bins is defined
within this region, with the central bin identical to the narrow
signal window. It is assumed that the
background consists of a combinatorial component, which is described
by a two-dimensional quadratic function, and a \Lcp component,
which is described by the product of a signal peak in \mLcp
and a quadratic function in $\delta m$.
Under this assumption, the background distribution can be fully
determined from the 24 sideband bins and hence its integral within
the signal box calculated. In this way the value of $N_B$
and the associated statistical uncertainty are determined. This method has the
advantage that it requires only minor assumptions about the background
distribution, given that part of that distribution
cannot be studied prior to unblinding.
It is adopted as the baseline approach for this reason.

The second method, used as a cross-check, imposes a narrow window 
on all candidates of
$2273 < \mLcp < 2303$\mevcc to reduce the problem to a
one-dimensional distribution in $\delta m$.
Based on studies of the \mLcp and $\delta m$ sidebands,
it is found that the
background can be described by a function of the form
\begin{equation}
  f(\delta m) = \left\{
    \begin{array}{ll}
      \phantom{a} L(\delta m; \mu, \sigma_L) & \delta m \leq \mu \\
      a L(\delta m; \mu, \sigma_R) & \delta m \geq \mu \\
    \end{array}
  \right.
  \label{eq:landau}
\end{equation}
where $L(\delta m; \mu, \sigma)$ is a Landau distribution,
$a$ is chosen such that
$L(\mu; \mu, \sigma_L) = a L(\mu; \mu, \sigma_R)$,
and $\mu$, $\sigma_L$, and $\sigma_R$ are free parameters.
The data are fitted with this function across the full range,
$0 < \delta m < 1500$\mevcc, excluding the signal window of
width 20\mevcc. The fit function is then integrated across the
signal window to give the expected background $N_B$.

\section{Efficiency ratio}
\label{sec:effic}

To measure $R$, it is necessary to
evaluate
the ratio of efficiencies
for the normalisation and signal modes, 
$\varepsilon_{\text{norm}} / \varepsilon_{\text{sig}}$.
The method used to evaluate this ratio is described below.
The
signal efficiency depends upon the mass and lifetime of
the \Xiccp, neither of which is known. To handle this, simulated
events are generated with $m(\Xiccp)=3500$\mevcc and
$\tau_{\Xiccp}=333$\fs and the efficiency ratio is evaluated at this
working point. The variation of the efficiency ratio as
a function of $\delta m$ and $\tau_{\Xiccp}$ relative to the working
point is then determined with a reweighting technique as discussed
in Sec.~\ref{sec:variation}. The kinematic distribution of
\Xiccp produced at the LHC is also unknown, but unlike the mass and
lifetime it cannot be described in a model-independent way with a
single additional parameter.
Instead, the upper limits 
are evaluated assuming
the distributions produced by the \genxicc model.

The efficiency ratio may be
factorised into several components as
\begin{equation}
  \frac{
    \varepsilon_{\text{norm}}
  }{
    \varepsilon_{\text{sig}}
  } = 
  \frac{ \varepsilon_{\text{norm}}^{\text{acc}} }{ \varepsilon_{\text{sig}}^{\text{acc}} } \,
  \frac{ \varepsilon_{\text{norm}}^{\text{sel}|\text{acc}} }{ \varepsilon_{\text{sig}}^{\text{sel}|\text{acc}} } \,
  \frac{ \varepsilon_{\text{norm}}^{\text{PID}|\text{sel}} }{ \varepsilon_{\text{sig}}^{\text{PID}|\text{sel}} } \,
  \frac{1}{\varepsilon_{\text{sig}}^{\text{ANN}|\text{PID}}} \,
  \frac{ \varepsilon_{\text{norm}}^{\text{trig}|\text{PID}} }{ \varepsilon_{\text{sig}}^{\text{trig}|\text{ANN}} }
  ,
\end{equation}
where efficiencies are evaluated for
the acceptance (acc),
the reconstruction and selection excluding PID and the ANN (sel),
the particle identification cuts (PID),
the ANN selector (ANN) for the signal mode only,
and the trigger (trig).
Each element is the efficiency relative to all previous steps in the
order given above. 

In most cases the individual ratios are evaluated with
simulated \Xiccp and \Lcp decays, taking the fraction of
candidates that passed the requirement in question. However, in some
cases the efficiencies need to be corrected for known differences
between simulation and data. This applies to the efficiencies for
tracking, for passing PID requirements, and for passing the
calorimeter hardware trigger. Control samples of data are used to determine
these corrections as a function of track kinematics and event charged track
multiplicity, and the simulated events are weighted accordingly.
The data samples used are $J/\psi \to \mu^+ \mu^-$ for the tracking
efficiency, and $\Dstarp \to \Dz (\to \Km \pip) \pip$ and
$\varLambda \to p \pim$ for both the PID and calorimeter hardware
trigger requirements.
The track multiplicity distribution is taken from data for the \Lcp
sample, but for \Xiccp events it is not known. It is modelled by taking a
sample of events containing a reconstructed \Bs decay,
on the grounds that \Bs production also requires two non-light
quark-antiquark pairs.

The efficiency ratio obtained at this working point
is $\varepsilon_{\text{norm}} / \varepsilon_{\text{sig}} = 20.4$.
Together with the value for $N_{\text{norm}}$
obtained in Sec.~\ref{sec:yield:sig}
and the definition in eq.~\ref{eq:defAlpha},
this implies the single-event sensitivity $\alpha$ is $2.5 \times 10^{-5}$
at $m(\Xiccp)=3500$\mevcc, $\tau_{\Xiccp}=333$\fs.

\section{Systematic uncertainties}
\label{sec:sys}

The statistical uncertainty on the measured signal yield is the dominant uncertainty
in this analysis,
  and the systematic uncertainties on $\alpha$ have very limited effect
  on the expected upper limits.
  As in the previous section, they will be evaluated at the working point of
  $m(\Xiccp)=3500$\mevcc and $\tau_{\Xiccp}=333$\fs, and their variation
  with mass and lifetime hypothesis considered separately.
Of the systematic uncertainties, the largest ($18.0\%$) is due to the
limited sample size of simulated events used to calculate
the efficiency ratio. Beyond this, there are several instances where the
simulation may not describe the signal accurately in data. These are
corrected with control samples of data, with
systematic uncertainties, outlined below, assigned to reflect uncertainties
in these corrections.

The IP resolution of tracks in the VELO is found to be worse in data
than in simulated events. To estimate the impact of this effect on the signal efficiency,
a test is performed with simulated events in which the VELO resolution is artificially degraded to the same level.
This is found to change the efficiency
of the reconstruction and non-ANN selection by $6.6\%$, and that of
the ANN by $6.7\%$. Taking these effects to be fully correlated,
a systematic uncertainty of $13.3\%$ is assigned.

A track-by-track correction is applied to the PID efficiency based on
control samples of data. There are several systematic uncertainties
associated with this correction. The first is due to the
limited size of the control samples, notably for high-\pt protons
from the $\varLambda$ sample. The second is due to the
assumption that the corrections factorise between the tracks, whereas
in practice there are kinematic correlations. The third is due to the
dependence on the event track multiplicity. The fourth is due to
limitations in the method (\eg the finite kinematic binning used)
and is assessed by applying it to samples of
simulated events. The sum in quadrature of the above gives an
uncertainty of $11.8\%$.

Systematic uncertainties also arise from the tracking efficiency ($4.7\%$)
and from the hardware trigger efficiency ($3.3\%$). 
Additional systematic uncertainties associated with candidate multiplicity, yield
measurement, and the decay model of \XiccpToLcpKmpip,
which may proceed through intermediate resonances,
were considered but found to be negligible in comparison
with the total systematic uncertainty.
The systematic uncertainties
are summarised in Table~\ref{tab:sys}. Taking their sum in quadrature,
the total systematic uncertainty is $26\%$.

\begin{table}
  \caption{
    \small
    Systematic uncertainties on the single-event sensitivity $\alpha$.
  }
  \begin{center}
    \begin{tabular}[c]{lr}
      \hline
      \multicolumn{1}{c}{Source}   &     \multicolumn{1}{c}{Size} \\
      \hline
      Simulated sample size   &     $18.0\%$  \\
      IP resolution           &     $13.3\%$  \\
      PID calibration         &     $11.8\%$  \\
      Tracking efficiency     &     $ 4.7\%$    \\
      Trigger efficiency      &     $ 3.3\%$   \\
      \hline
      Total uncertainty       &     $26.0\%$  \\
      \hline
    \end{tabular}
  \end{center}
  \label{tab:sys}
\end{table}

\section{Variation of efficiency with mass and lifetime}
\label{sec:variation}

The efficiency to trigger on, reconstruct, and select \Xiccp
candidates has a strong dependence upon the \Xiccp lifetime.
The efficiency also depends upon the \Xiccp mass, since
this affects the opening angles and the \pt of the daughters.

The simulated \Xiccp events are generated with a proper decay time
distribution given by an exponential function of average lifetime
$\tau_{\Xiccp}=333$\fs. To test other lifetime hypotheses,
the simulated events are reweighted to follow a different
exponential distribution and the efficiency is recomputed.
Most systematic uncertainties are unaffected, but those
associated with the limited simulated sample size and with the
hardware trigger efficiency increase at shorter lifetimes
(the latter due to kinematic correlations rather than
direct dependence on the decay time distribution).
The values and uncertainties of the single-event
sensitivity $\alpha$ are given for several lifetime
hypotheses in Table~\ref{tab:varyLifetime}.

\begin{table}
  \caption{
    \small
    Single-event sensitivity $\alpha$ for different lifetime
    hypotheses $\tau$, assuming $m(\Xiccp)=3500$\mevcc.
    The uncertainties quoted include statistical and systematic
    effects, and are correlated between different lifetime
    hypotheses.
  }
  \begin{center}
    \begin{tabular}{cd{1}@{\,$\pm$\,}d{1}}
      \hline
      $\tau$   &  \multicolumn{2}{c}{$\alpha$\,($\times 10^{-5}$)} \\
      \hline
      100\fs    &   63   & 31 \\
      150\fs    &   15   &  5 \\
      250\fs    &    4.1 &  1.1 \\
      333\fs    &    2.5 &  0.6 \\
      400\fs    &    1.9 &  0.5 \\
      \hline
    \end{tabular}
  \end{center}
  \label{tab:varyLifetime}
\end{table}

To assess the effect of varying the \Xiccp mass hypothesis,
large samples of simulated events are generated 
for two other mass hypotheses, 
$m(\Xiccp)=3300$\mevcc
and 3700\mevcc,
without running the \geant detector simulation.
Two tests are carried out with these samples. First, the
detector acceptance efficiency is recalculated.
Second, the \pt distributions of the three daughters
of the \Xiccp in the main $m(\Xiccp)=3500$\mevcc sample
are reweighted to match those seen at
the other mass hypotheses and the remainder of the
efficiency is recalculated. 
  In both cases the systematic uncertainties are also recalculated,
  though very little change is found.
Significant variations in
individual components of the efficiency are
seen---notably in the acceptance, reconstruction, non-ANN
selection, and hardware trigger efficiencies---but when combined
cancel almost entirely.
  This is shown in Table~\ref{tab:varyMass}, which gives the
  value of $\alpha$ including the mass-dependent effects discussed
  above but excluding the correction for the efficiency of the $\delta m$
  signal window described in Sec.~\ref{sec:yield:sig} ($\alpha_u$),
  the correction for the variation in resolution,
  and the combined value of $\alpha$.
  Because the variation of $\alpha_u$ with mass is extremely small,
  a simple first-order correction is sufficient.
  A straight line is fitted to the three points
  in the table and used to interpolate the fractional
  variation in $\alpha_u$ between the mass hypotheses.
  The resolution correction is then applied separately.
  Due to the smallness of the mass-dependence,
  correlations between variation with mass and with
  lifetime are neglected.

\begin{table}
  \caption{
    \small
    Variation in single-event sensitivity for different mass
    hypotheses $m(\Xiccp)$, assuming $\tau=333$\fs.
    The uncertainties quoted include statistical and systematic
    effects, and are correlated between different mass
    hypotheses.
      The variation is shown separately for all effects other
      than the efficiency of the $\delta m$ window ($\alpha_u$), 
      for the correction due to the mass-dependent resolution,
      and for the combination ($\alpha$).
  }
  \begin{center}
    \begin{tabular}{cd{2}@{\,$\pm$\,}d{2}cd{2}@{\,$\pm$\,}d{2}}
      \hline
       $m(\Xiccp)$  &  \multicolumn{2}{c}{$\alpha_u$\,($\times 10^{-5}$)} & Resolution correction & \multicolumn{2}{c}{$\alpha$\,($\times 10^{-5}$)}\\
      \hline
      3300\mevcc & 2.29 & 0.61 & 0.992 & 2.30 & 0.62 \\
      3500\mevcc & 2.38 & 0.62 & 0.957 & 2.49 & 0.65 \\
      3700\mevcc & 2.36 & 0.63 & 0.903 & 2.61 & 0.70 \\
      \hline
    \end{tabular}
  \end{center}
  \label{tab:varyMass}
\end{table}

  As explained in Sec.~\ref{sec:intro}, the value of $R$ at LHCb is not
  well known but is expected to be of the order $10^{-5}$ to $10^{-4}$,
  while the SELEX observation corresponds to $R=9\%$.
  Table~\ref{tab:expected} shows the expected signal yield,
  calculated according to eq.~\ref{eq:defineMainResult},
  for various values of $R$ and lifetime hypotheses.
  From studies of the sidebands in \mLcp and $\delta m$,
  the expected background in the narrow signal window is
  between 10 and 20 events. Thus, no significant signal excess is expected
  if the value of $R$ at LHCb is in the range suggested by theory.
  However, if production is greatly enhanced for baryon-baryon
  collisions at high rapidity, as reported at SELEX,
  a large signal may be visible.
  The procedure for determining the significance of a signal, or for
  establishing limits on $R$, is discussed in the following section.

\begin{table}
  \caption{
    \small
      Expected value of the signal yield $N_{\text{sig}}$
      for different values of $R$ and lifetime hypotheses,
      assuming $m(\Xiccp)=3500$\mevcc.
      The uncertainties quoted are due to the systematic uncertainty
      on $\alpha$.
  }
  \begin{center}
    \begin{tabular}{cd{0}@{\,$\pm$\,}d{0}d{1}@{\,$\pm$\,}d{1}d{2}@{\,$\pm$\,}d{2}}
      \hline
      $\tau$ & \multicolumn{2}{c}{$R=9\%$} & \multicolumn{2}{c}{$R=10^{-4}$} & \multicolumn{2}{c}{$R=10^{-5}$} \\
      \hline
      100\fs & 140 & 70 & 0.2 & 0.1 & 0.02 & 0.01 \\
      150\fs & 600 & 200 & 0.7 & 0.2 & 0.07 & 0.02 \\
      250\fs & 2200 & 600 & 2.4 & 0.7 & 0.24 & 0.07 \\
      333\fs & 3600 & 900 & 4.0 & 1.0 & 0.40 & 0.10 \\
      400\fs & 4800 & 1200 & 5.3 & 1.4 & 0.53 & 0.14 \\
      \hline
    \end{tabular}
  \end{center}
  \label{tab:expected}
\end{table}

\section{Tests for statistical significance and upper limit calculation}
\label{sec:UL}

Since $m(\Xiccp)$ is \textit{a priori} unknown, tests
for the presence of a signal are carried out at numerous mass hypotheses,
between $\delta m = 380$\mevcc and $\delta m = 880$\mevcc inclusive
in 1\mevcc steps for a total of 501~tests.
For a given value of $\delta m$, the signal and background
yields and their associated statistical uncertainties are estimated as described in
Sec.~\ref{sec:yield:sig}. From these
the local significance $\mathscr{S}(\delta m)$ is calculated, where
$\mathscr{S}(\delta m)$ is defined as
\begin{equation}
  \mathscr{S}(\delta m) \equiv \frac{N_{S+B} - N_B}{\sqrt{\sigma_{S+B}^2 + \sigma_{B}^2}}
  \label{eq:signif}
\end{equation}
and $\sigma_{S+B}$ and $\sigma_{B}$ are the estimated
statistical uncertainties on the yield in the signal window
and on the expected background, respectively. Since
multiple points are sampled, the
look elsewhere effect (LEE)~\cite{Lyons:1900zz} 
must be taken into account.
The procedure used is to generate a large number of
pseudo-experiments containing only background events, with the amount
and distribution of background chosen to match the data (as
estimated from sidebands). For each pseudo-experiment, the full
analysis procedure is applied in the same way as for data, and the
local significance is measured at all 501 values of $\delta m$.
The LEE-corrected \pvalue for a given $\mathscr{S}$ is then
taken to be the fraction of the pseudo-experiments that
contain an equal or larger local significance at any point
in the $\delta m$ range.

The procedure established before unblinding is that
if no signal with an LEE-corrected significance of at least $3\sigma$ 
is seen, upper limits on $R$ will be quoted.
The $CL_s$ method~\cite{Read:2002hq,Cowan:2010js}
is applied to determine upper limits on $R$ for a
particular $\delta m$ and lifetime hypothesis,
given the observed yield $N_{S+B}$ and expected
background $N_B$ in the signal window obtained
as described in Sec.~\ref{sec:yield:sig}.
The statistical uncertainty on $N_B$ and
systematic uncertainties on $\alpha$ are 
taken into account.
The 95\% CL upper limit is then taken as the value of
$R$ for which $CL_s = 0.05$. Upper limits are calculated
at each of the 501 $\delta m$ hypotheses, and for five lifetime
hypotheses (100, 150, 250, 333, 400\fs).

\section{Results}
\label{sec:res}

\begin{figure}[htbf]
  \begin{center}
    \includegraphics[width=0.49\textwidth]{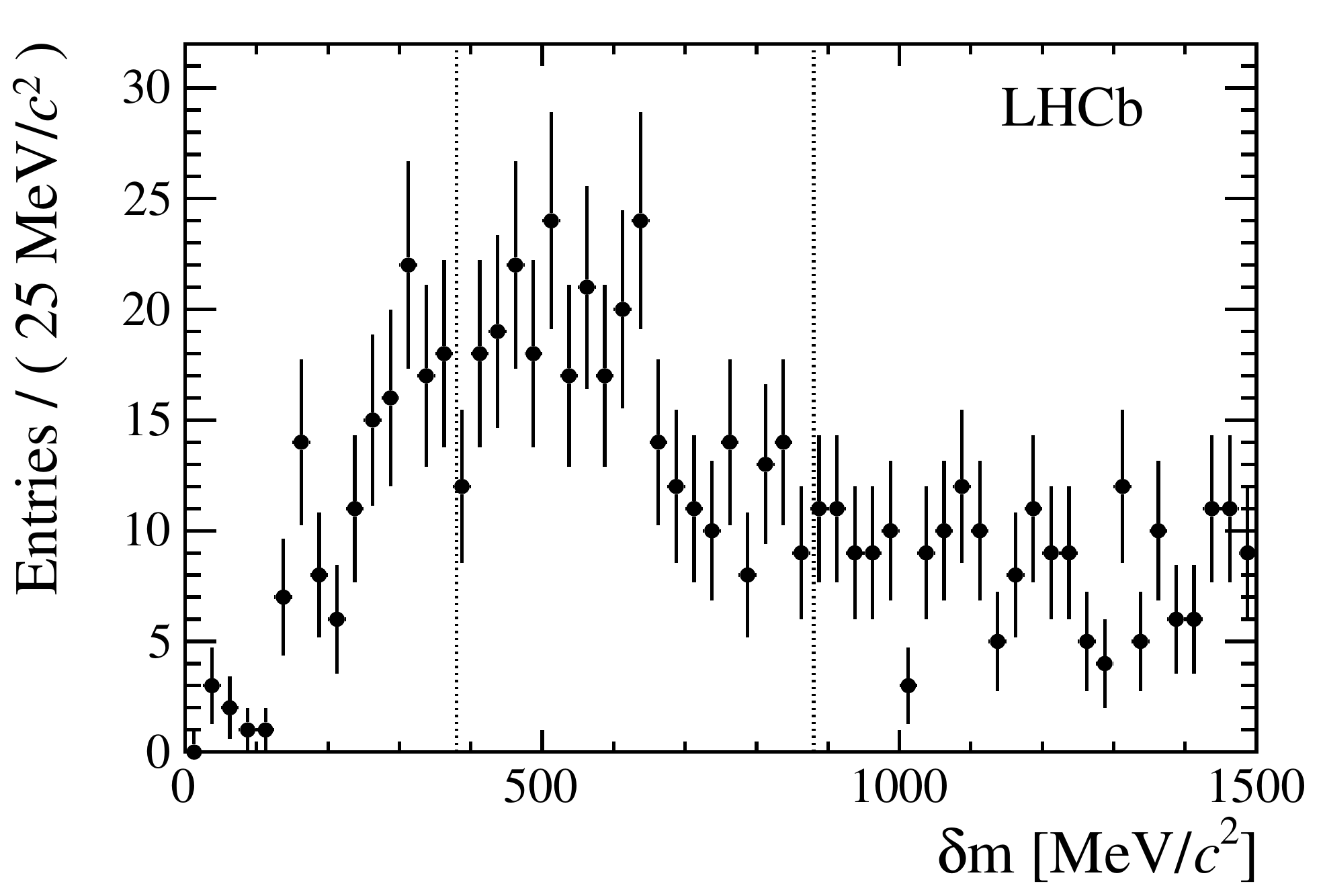}
    \includegraphics[width=0.49\textwidth]{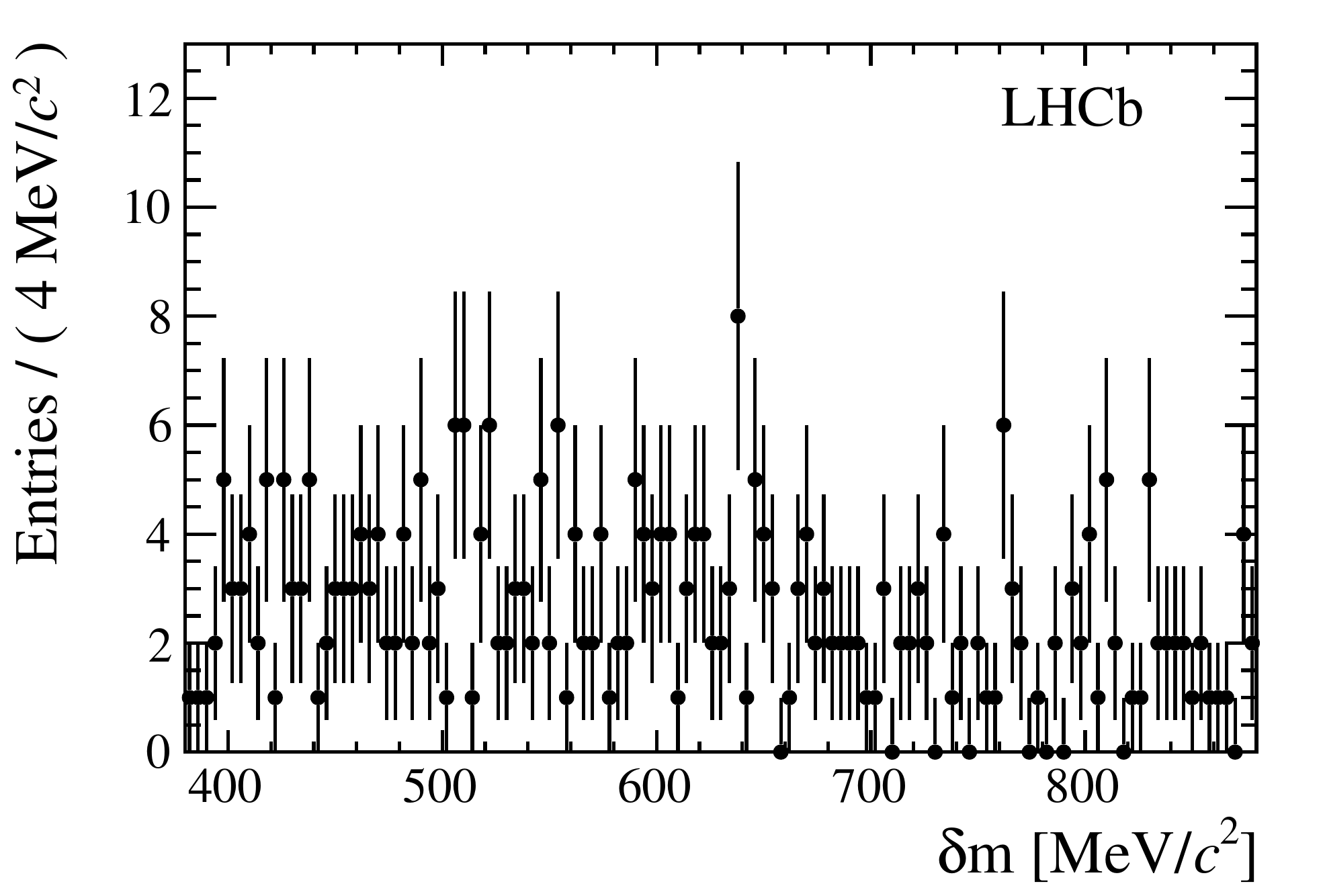}
  \end{center}
  \caption{
    \small
    Spectrum of $\delta m$ requiring $2273 < \mLcp < 2303$\mevcc.
    Both plots show the same data sample, but with different $\delta m$
    ranges and binnings. The wide signal region is shown in the right
    plot and indicated by the dotted vertical lines in the left plot.
  }
  \label{fig:unblindRawSpectra}
\end{figure}

\begin{figure}[htbf]
  \begin{center}
    \includegraphics[width=0.49\textwidth]{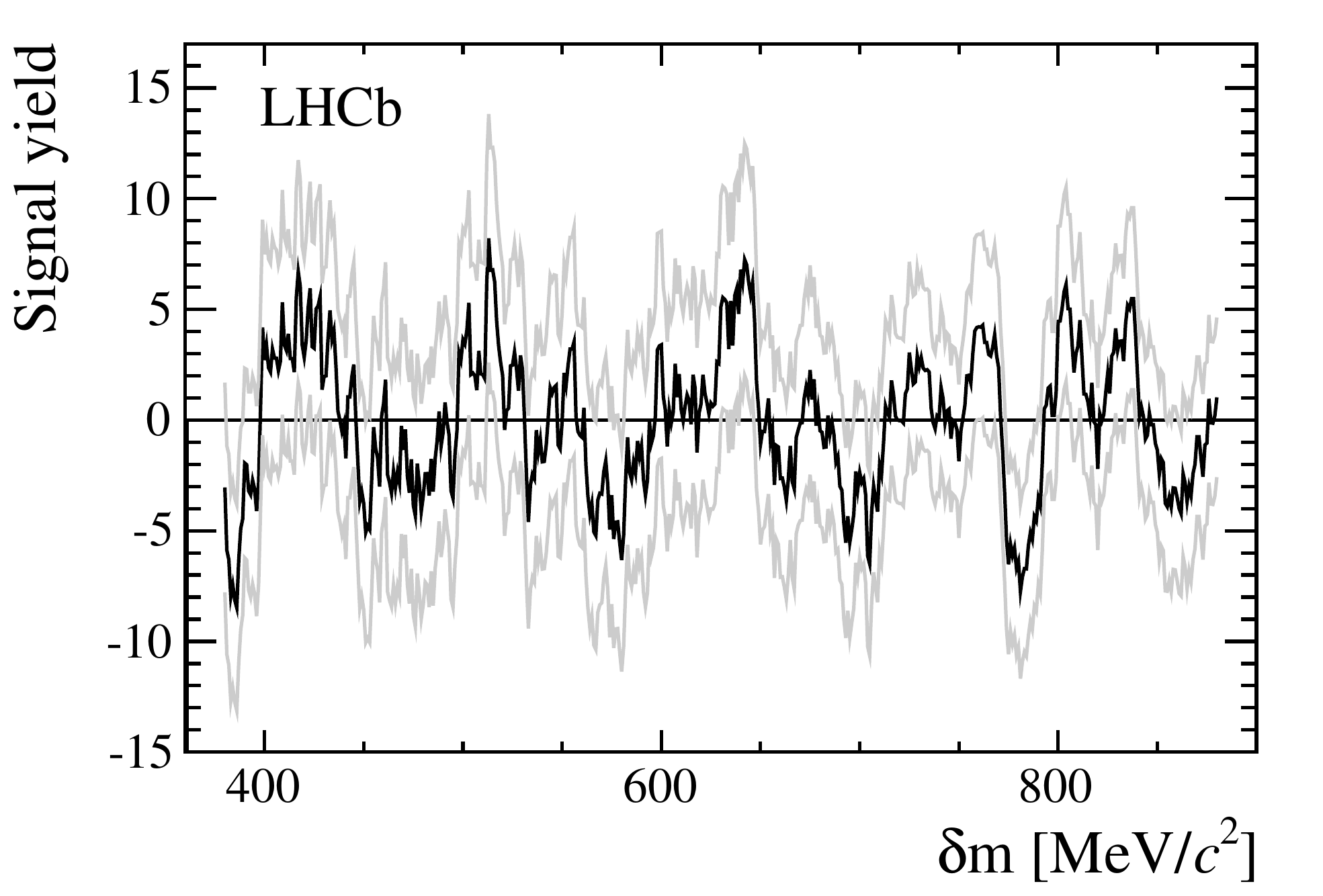}
    \includegraphics[width=0.49\textwidth]{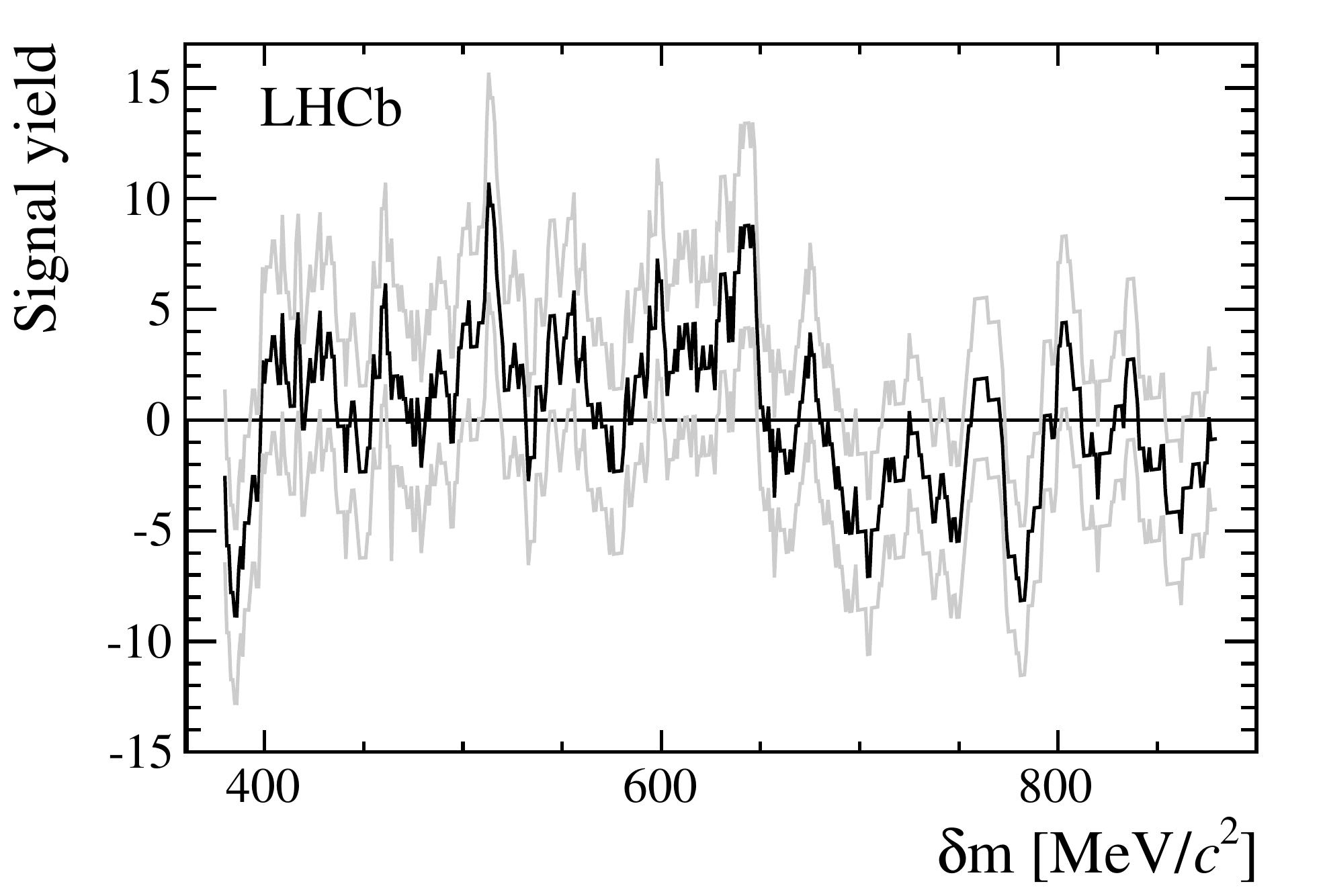} \\
    \includegraphics[width=0.75\textwidth]{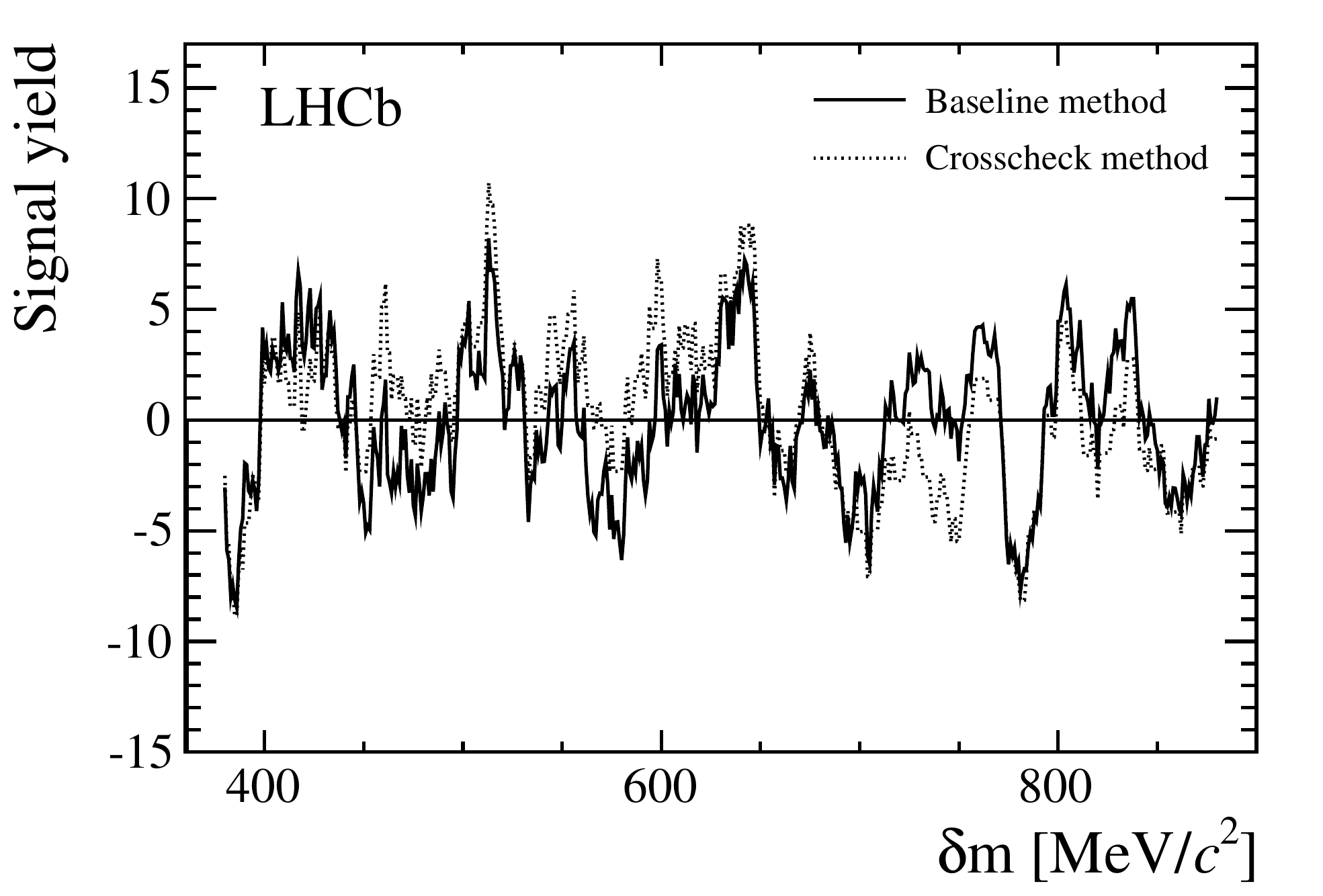}
  \end{center}
  \caption{
    \small
    Measured signal yields as a function of $\delta m$.
    The upper two plots show the estimated signal yield as a dark line and the
    $\pm 1\sigma$ statistical error bands as light grey lines for
    (upper left)~the baseline method and
    (upper right)~the cross-check method. 
    The central values of the two methods
    are compared in the lower plot and found to agree well.
  }
  \label{fig:unblindYields}
\end{figure}

\begin{figure}[htbf]
  \begin{center}
    \includegraphics[width=0.7\textwidth]{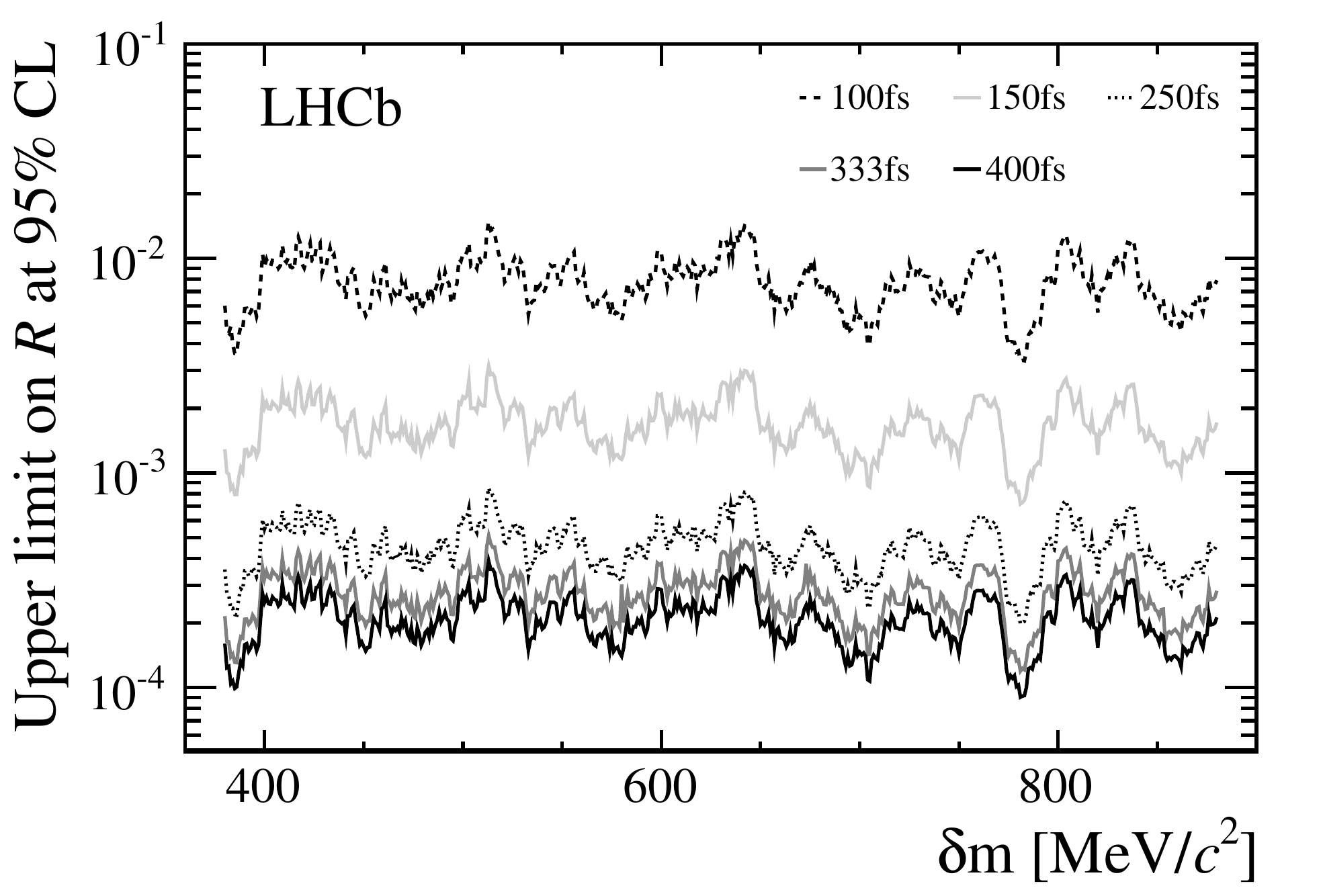}
  \end{center}
  \caption{
    \small
    Upper limits on $R$ at the 95\% CL as a function of $\delta m$,
    for five \Xiccp lifetime hypotheses.
  }
  \label{fig:unblindUL}
\end{figure}

The $\delta m$ spectrum in data is shown in Fig.~\ref{fig:unblindRawSpectra},
and the estimated signal yield in Fig.~\ref{fig:unblindYields}.
No clear signal is found with either background subtraction method. In both cases the
largest local significance occurs at $\delta m = 513$\mevcc, with
$\mathscr{S} = 1.5\sigma$ in the baseline method and $\mathscr{S}=2.2\sigma$ in the cross-check.
Applying the LEE correction described in Sec.~\ref{sec:UL}, these
correspond to \pvalues of 99\% and 53\%, respectively. Thus, with
no significant excess found above background, upper limits are set on $R$
at the 95\% CL, shown in Fig.~\ref{fig:unblindUL} for the first
method. These limits are tabulated in Table~\ref{tab:unblindUL}
for blocks of $\delta m$ and the five lifetime hypotheses.
The blocks are 50\mevcc wide, and for each block the largest (worst)
upper limit seen for a $\delta m$ point in that block is given. Similarly,
the largest upper limit seen in the entire 500\mevcc mass range is also given.
A strong dependence in sensitivity on the lifetime hypothesis is seen.

\begin{table}[htbf]
  \caption{
    \small
    Largest values of the upper limits (UL) on $R$ at the 95\% CL
    in blocks of $\delta m$
    for a range of lifetime hypotheses,
    given in units of $10^{-3}$.
    The largest values across the entire 500\mevcc range are also shown.
  }
 \begin{center}
    \begin{tabular}{crcccc}
      \hline
       & \multicolumn{5}{c}{$R$, largest 95\% CL UL in range $\times 10^{3}$} \\
$\delta m$ (MeV$/c^{2}$) & \multicolumn{1}{c}{100\fs} & 150\fs& 250\fs& 333\fs& 400\fs \\ \hline
380--429 & 12.6 & 2.7 & 0.73 & 0.43 & 0.33 \\
430--479 & 11.2 & 2.4 & 0.65 & 0.39 & 0.29 \\
480--529 & 14.8 & 3.2 & 0.85 & 0.51 & 0.39 \\
530--579 & 10.7 & 2.3 & 0.63 & 0.38 & 0.29 \\
580--629 & 10.9 & 2.3 & 0.63 & 0.38 & 0.29 \\
630--679 & 14.2 & 3.0 & 0.81 & 0.49 & 0.37 \\
680--729 &  9.5 & 2.0 & 0.56 & 0.33 & 0.25 \\
730--779 & 10.8 & 2.3 & 0.63 & 0.37 & 0.28 \\
780--829 & 12.8 & 2.8 & 0.74 & 0.45 & 0.34 \\
830--880 & 12.2 & 2.6 & 0.70 & 0.42 & 0.32 \\
\hline
380--880 & 14.8 & 3.2 & 0.85 & 0.51 & 0.39 \\ \hline

    \end{tabular}
  \end{center}
  \label{tab:unblindUL}
\end{table}

\begin{figure}
  \begin{center}
    \includegraphics[width=0.45\textwidth]{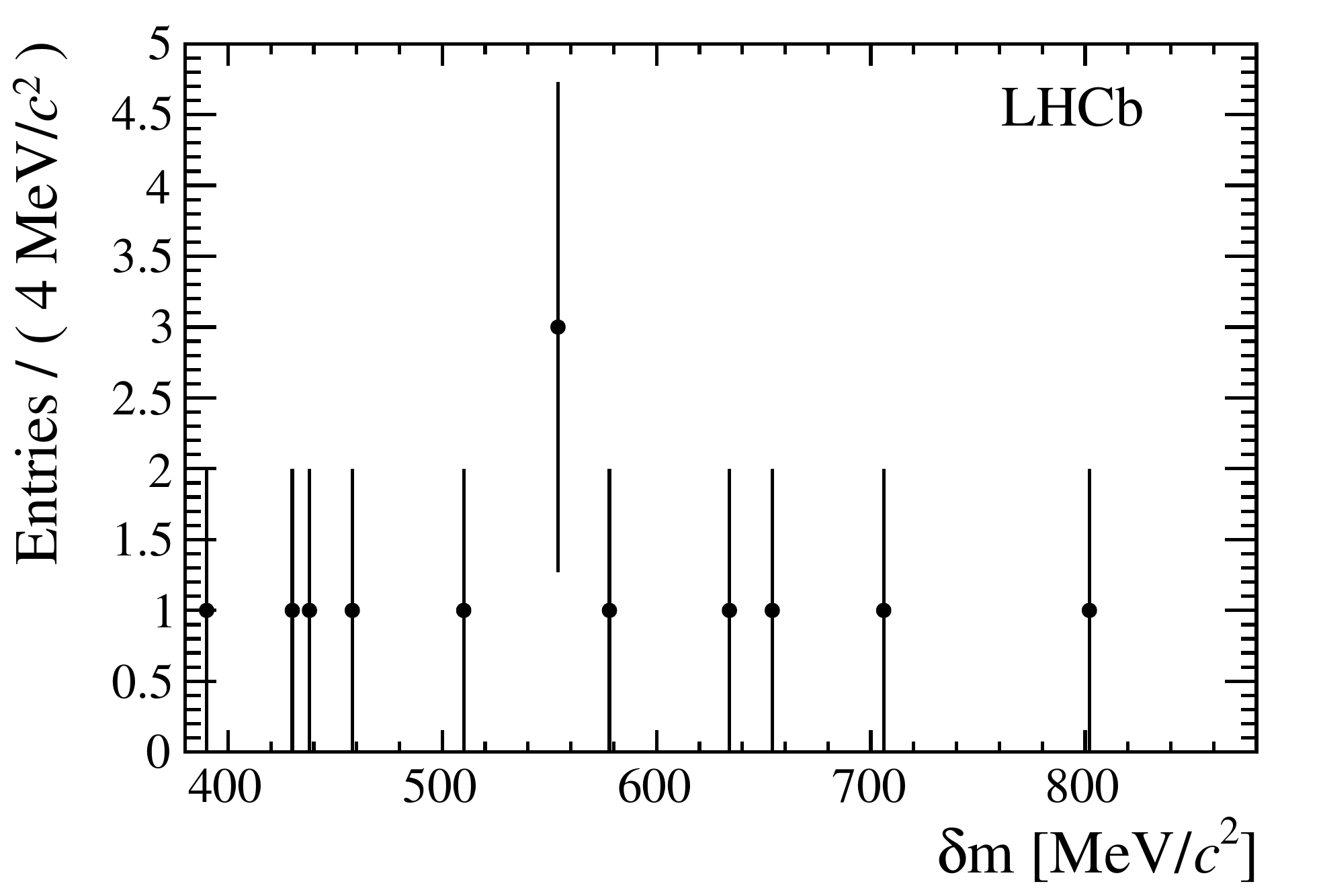}
    \includegraphics[width=0.45\textwidth]{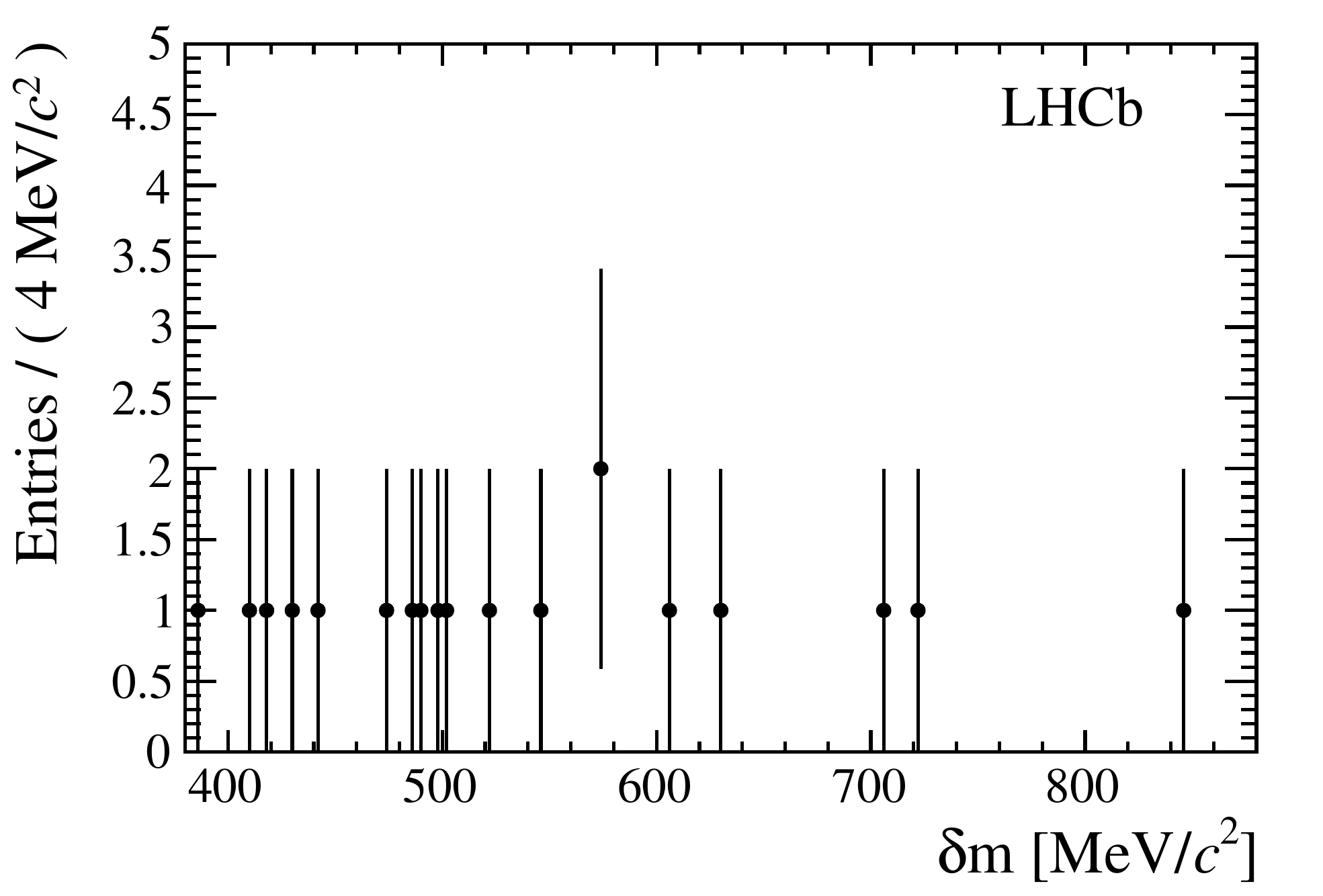}
  \end{center}
  \caption{
    \small
    Mass difference spectrum requiring $2273 < \mLcp < 2303$\mevcc.
    Candidates are also required to be consistent with
    (left)~an intermediate $\Sigma_{\cquark}(2455)^{++}$,
    (right)~an intermediate $\Sigma_{\cquark}(2520)^{++}$.
  }
  \label{fig:unblindSc}
\end{figure}

The decay \XiccpToLcpKmpip may proceed through an
intermediate $\Sigma_{\cquark}^{++}$ resonance. Such decays would be
included in the yields and limits already shown. Nonetheless,
further checks are made with an explicit requirement that the
$\Lcp \pip$ invariant mass be consistent with that of a $\Sigma_{\cquark}^{++}$, since this
substantially reduces the combinatorial background.
For $\Sigma_{\cquark}(2455)^{++}$ and $\Sigma_{\cquark}(2520)^{++}$,
the mass offsets $\left[ m([\proton \Km \pip]_{\Lc} \pip) - \mLcp \right]$
are required to be within 4\mevcc and 15\mevcc of the world-average value, respectively.
The resulting $\delta m$ spectra are shown
in Fig.~\ref{fig:unblindSc}. No statistically significant excess
is present.

\section{Conclusions}
\label{sec:conc}

A search for the decay \XiccpToLcpKmpip is performed
at LHCb
  with a data sample
  of $\proton\proton$ collisions,
  corresponding to an integrated luminosity of 0.65\invfb,
  recorded at a centre-of-mass energy of 7\tev.
No significant signal is found. Upper limits on 
the \Xiccp cross-section times branching fraction relative to the
\Lcp cross-section are obtained for a range of mass and lifetime
hypotheses, assuming that the kinematic distributions of the \Xiccp
follow those of the \genxicc model.
The upper limit depends strongly on the lifetime, varying from 
  $1.5 \times 10^{-2}$ for 100\fs to
  $3.9 \times 10^{-4}$ for 400\fs.
  These limits are significantly below the value of $R$ found at SELEX. This may be
  explained by the different production environment, or if the
  \Xiccp lifetime is indeed very short ($\ll 100$\fs).
Future searches at LHCb with
  improved trigger conditions,
  additional $\Xi_{cc}$ decay modes,
  and larger data samples
should improve the sensitivity significantly, especially at short lifetimes.

\section*{Acknowledgements}

\noindent We express our gratitude to our colleagues in the CERN
accelerator departments for the excellent performance of the LHC. We
thank the technical and administrative staff at the LHCb
institutes. We acknowledge support from CERN and from the national
agencies: CAPES, CNPq, FAPERJ and FINEP (Brazil); NSFC (China);
CNRS/IN2P3 and Region Auvergne (France); BMBF, DFG, HGF and MPG
(Germany); SFI (Ireland); INFN (Italy); FOM and NWO (The Netherlands);
SCSR (Poland); MEN/IFA (Romania); MinES, Rosatom, RFBR and NRC
``Kurchatov Institute'' (Russia); MinECo, XuntaGal and GENCAT (Spain);
SNSF and SER (Switzerland); NAS Ukraine (Ukraine); STFC (United
Kingdom); NSF (USA). We also acknowledge the support received from the
ERC under FP7. The Tier1 computing centres are supported by IN2P3
(France), KIT and BMBF (Germany), INFN (Italy), NWO and SURF (The
Netherlands), PIC (Spain), GridPP (United Kingdom). We are thankful
for the computing resources put at our disposal by Yandex LLC
(Russia), as well as to the communities behind the multiple open
source software packages that we depend on.


\addcontentsline{toc}{section}{References}
\setboolean{inbibliography}{true}
\bibliographystyle{LHCb}
\bibliography{main,Xicc,LHCb-PAPER,LHCb-CONF,LHCb-DP}

\end{document}